# Ads/CFT correspondence in condensed matter

A.S.T. Pires

Universidade Federal de Minas Gerais, Departamento de Fisica, Belo Horizonte, MG, Brazil.

e-mail: antpires@fisica.ufmg.br

**Abstract**

The goal of these notes is to introduce, in a very elementary way, the idea of the anti de-Sitter/Conformal Field Theory (AdS/CFT) correspondence to condensed matter physicists. This theory relates a gravity theory in a (d+1)- dimensional anti-de Sitter spacetime to a strongly-coupled d- dimensional quantum field theory living on its boundary. The AdS/CFT correspondence can be used to study finite temperature real time processes, such as response functions and dynamics far from equilibrium in quantum critical points in condensed matter systems. Computation of these quantities is reduced to solving classical gravitational equations in one higher dimension than the original theory.

## 1. Introduction

The so called AdS/CFT (anti-de Sitter/ Conformal field theory) correspondence [1-3] have become very important in higher energy theoretical physics over the last ten years, and hundreds of papers have being published about the subject. One reason is that this correspondence can be used to understand strongly interacting field theories by mapping them to classical gravity. The main idea is as follows: Large *N* gauge theories (which are strongly coupled conformal field theories) in *d* space-time dimensions are mapped to a classical gravitational theory in $d$ + 1 space-time dimensions which are asymptotically AdS. In fact, it is a strong-weak coupling duality: when one theory is coupled weakly, the dual description involves strong coupling, and vice versa. Different CFTs will correspond to theories of gravity with different field content and different bulk actions, e.g. different values of the coupling constants in the bulk. The AdS/CFT correspondence is an example of a more general technique called “the holographic principle”, which was motivated by the study of the thermodynamic properties of a black hole. As it is well known, the entropy of a black hole is proportional to the area of its event horizon. Thus, the number of degrees of freedom needed to describe a quantum black hole scales with its area, not its volume. We can interpret this result saying that the physics inside the black hole is mapped onto its horizon, some kind of a “hologram”.

Arguments showing that some quantum field theories (QPT) are secretly quantum theories of gravity and therefore we can use them to compute observables of the QFT, when the gravity theory is classical, can be found in [4]. The number of theories that can be studied using the correspondence is still small, and it does not include any theory that describes a known physical system. However it is hoped that these theories can be used to capture essential features of theories realized in nature [1].

Recently, several authors have claimed that some phenomena in condensed matter systems, such as quantum phase transitions, also provide a candidate for the use of AdS/CFT [5-15].

In these notes I intend to introduce to condensed matter physicists the AdS/CFT correspondence presenting just the general ideas. The interested reader can consult the references to get more information. Nothing original is presented here and I have used material from other review articles including [4,16-18]. I have presented only relevant materials, omitting the details. In Sec. 2, I introduce quantum phase transitions for the physicists with no knowledge in the subject. For more information I recommend the excellent book by Sachdev[19]. For more recent materials see his web page. In Sec.3, it is shown that choosing a metric, in a manifold with an extra dimension, with all the symmetries of the field theory at the critical point, leads to a metric of a space which is AdS. The theory is however at zero temperature. To work at finite temperature a black hole is introduced in the theory in Sec.4. In condensed matter physics we are generally interested in the presence of an electric or magnetic field. This is treated in Sec.5. In Sec. 6, I present a brief introduction to the linear response theory. In Sec.7 and 8 I talk about transport and in Sec.9 I discuss briefly spin transport in the model presented in the section below. In appendix A, I list some equations of the general relativity theory, so that the reader can perform calculations without knowing the subject, just following a receipt. Finally in Appendix B, I introduce the concept of anti-de Sitter space.

## 2. Quantum phase transitions

Quantum phase transitions (QPT) occur at zero temperature when a nonthermal parameter, let us say *g*, like pressure, chemical composition or magnetic field is varied [20,21]. The effects of quantum criticality, different from the classical case, might be observed at high temperatures. This is, in many cases, the signature of a quantum critical point might extend to high temperatures. One model where QPT can be well understood is the two-dimensional anisotropic quantum XY model described by the following Hamiltonian:

$$H = J\sum_{<n,m>}(S_n^x S_m^x + S_n^y S_m^y) + D\sum_n (S_n^z)^2, \tag{1}$$

where <n,m> represents the sum over nearest neighbors on the sites, n, of a regular lattice and $0 \le D < \infty$, where $D$ represents an *xy* easy-plane single ion anisotropy. Due to the form of the single-ion anisotropy we should have S > ½ and so we take S = 1. For $D$ less than a critical value $D_C$, the system has a thermal phase transition at a temperature $T_{BKT}$, the Berezinskii-Kosterlitz-Thouless temperature. This is another reason why this model is so interesting to be studied. This phase transition is associated with the emergence of a topological order, resulting from the pairing of vortices with opposite circulation. The BKT mechanism does not involve any spontaneous symmetry-breaking and emergence of a spatially uniform order parameter. The low-temperature phase is associated with a quasi-long-range order, at finite temperature, with the correlation of the order parameter decaying algebraically in space. Above the critical temperature the correlation decays exponentially.

For strong planar anisotropy in Hamiltonian (1) we have the so called large-$D$ phase. This phase consists of a unique ground state with total magnetization $S^z_{total} = 0$ separated by a gap from the first excited states which lie in the sectors $S^z_{total} = \pm 1$. The elementary excitations are *excitons*, with S = 1 and an infinite lifetime at low energies. For small $D$ the Hamiltonian (1) is in a gapless phase well described by the spin wave formalism. The increase of the anisotropy parameter $D$ reduces the transition temperature and at $D_C$ the critical temperature vanishes. Thus, at $D_C$ the system undergoes a QPT, at $T = 0$, from a gapless to a gaped phase.

The phase diagram for Hamiltonian (1) is expected to be as shown qualitatively in Fig. 1. The solid line in Fig.1 represents the line of critical points, determined by the BKT transition that terminates at the critical point $D_C$. Below this line, the inverse correlation length, $\xi^{-1}$, vanishes. The dashed line represents a crossover from the quantum critical region (where $\xi^{-1} \propto T$), to a region where quantum fluctuations dominate ($\xi^{-1} >> T$).

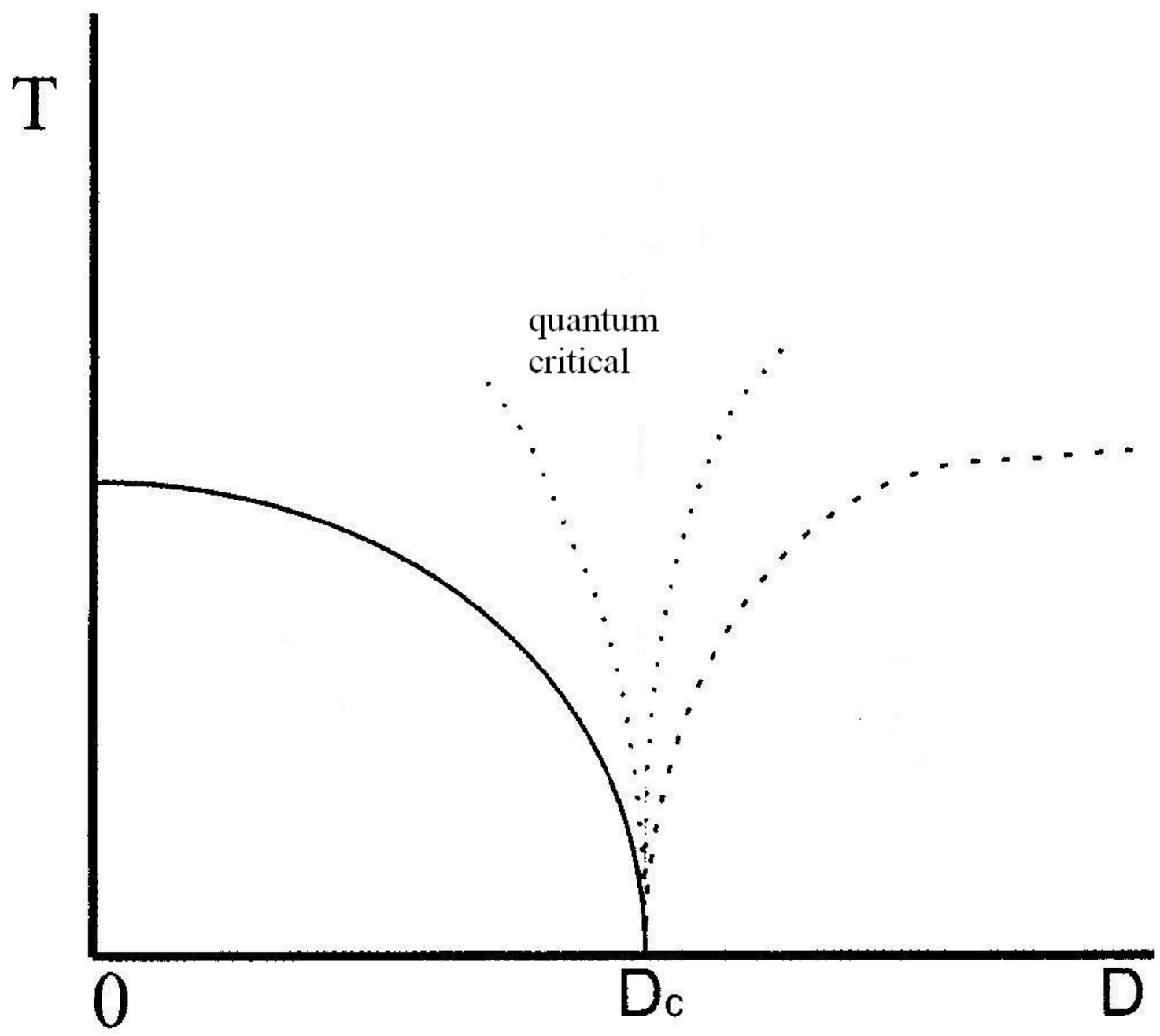

Figure 1: Phase diagram of the model described by the Hamiltonian (1).

At small $D$ ($D < D_C$) and across the BKT transition, we can use a self-consistent harmonic approximation to describe the model. The large $D$ phase can studied using the bond operator technique. In the intermediate quantum-critical regime neither description is adequate. A lack of well-defined weakly coupled quasiparticles in this region makes the description of quantum criticality difficult to study using traditional methods. Outside of AdS/CFT there are no models of strongly coupled quantum criticality in 2+1 dimensions in which analytic results for processes such as transport can be obtained [16].

In the low temperature phase of the region $D < D_C$ we have a power-law decay of the correlations but no broken symmetry. This phase is not described by a simple order parameter in the Landau-Ginzburg (LG) theory of phase transitions. We can, however, use the LG formalism to study the quantum-phase transition if for $D < D_C$ we restrict ourselves only to the $T = 0$ limit. The low energy dynamics of the model is described by the action [19]:

$$F = \int d^d r d\tau \frac{1}{2}\left[ c^2 (\nabla\phi)^2 + (\partial_\tau \phi)^2 + \delta\phi + \frac{V}{2}\phi^4 \right] , \quad (2)$$

where here $\phi$ is a field variable and τ is an imaginary time (τ = *it*). The transition is now tuned by varying $\delta \propto (D - D_C)$. The system has also Lorentz symmetry, although here $c$ is not the speed of light.

Other example of QPT can be found in [16] and [19]. I will mention just a few here. Néel- VBS (valence-bond solid) transition in antiferromagnets, superconducting-insulator transitions in thin metallic films, QPT between strongly Fermi-liquids as found in heavy fermions intermetallics and high $T_c$ superconductors and a change from electron to hole character as a function of bias voltage in graphene.

## 3. AdS/CFT correspondence

At the fixed point, the theory describing a QPT is invariant under rotations, translations, time translations and dilatations. The theory is also scale invariant. This means that the transformation $x^\mu \to \lambda x^\mu$ $(\mu = 0,1,2,d-1)$ is a symmetry. (The more general case $t \to \lambda^z t$, where $z$ is the dynamical critical exponent, will not be considered here).

Now we should look for a metric, in one higher dimension than the field theory, in which these symmetries are realized. Note that, to have translational invariance in $t$ and $x^i$ directions, we have to assume that all metric components depend on the extra dimension only. One metric with these properties is given by:

$$ds^2 = \frac{L^2}{r^2}\left(-dt^2 + dr^2 + dx^i dx^i\right). \quad (3)$$

The coordinates $\{t, x^i\}$ parameterize the space on which the field theory lives, while $r$ is the extra coordinate running from $r = 0$ (the ′boundary`) to $r = \infty$ (the ′horizon`). This is the metric for the anti-de Sitter space- time (see appendix B). (For general $z$, the term in $dt^2$ in the metric is written as $- dt^2/r^{2z}$). The indices $\{\mu,\nu\}$ will run over d spacetime dimensions of the QFT, $\{i, j\}$ over the spatial coordinates $x^i$, and {A,B} run over the full $d + 1$ dimensional bulk and the parameter $L$ sets the radius of curvature of the AdS space-time. The AdS manifold is a collection of copies of d-dimensional Minkowski space of varying size (constant $r$ slices are just copies of Minkowski space). Of course this metric can be written in different forms using different coordinates and in the above equation we have used the so called Poincaré coordinates.

The metric (3) enjoys also Lorentz boost symmetry and special conformal symmetries. It is believed that the classical dynamics about this background metric describes the physics of the strongly coupled theory.

As pointed out by Mc Greevy [4] the extra dimension of space time has a clear physical meaning. The renormalization group (RN) equations for the behavior of the coupling constants, in the field theory, as a function of the RG scale $u$ are [22]

$$u\frac{\partial g}{\partial u} = \beta[g(u)], \tag{4}$$

where the symbol β is used for historical reasons. This suggests that we can take the extra dimension as an energy scale. At the critical point β = 0 and the system is self-similar. Dimensional analysis says that $u$ will scale under the scale transformation as $u \to u/\lambda$. We can do a change of coordinate $r \equiv L^2/u$ and arrive at the metric in (3).

Taking the metric (3) into the Einstein equation of motion (A.8) we obtain $\Lambda = -d(d+1)/2L^2$. Inserting this relation into the Einstein's equation for an empty space and a negative cosmological constant, Eq. (A.10), and remembering that $n = d+1$, we arrive at

$$R_{ab} = -\frac{d}{L^2}g_{ab}. \tag{5}$$

Maldacena [1] conjectured the AdS/CFT duality, but his conjecture did not specify the precise way in which these two theories should be mapped onto each other. Later a proposal was made by Gubser, Klebanov and Polyakov [2] and by Witten[3]. The formulation is as follows.

Suppose that a field $\phi$ in the bulk is coupled to an operator $O$ on the boundary, such that the interaction Lagrangian is $\phi O$. The map between AdS and CFT quantities is given by

$$\left\langle \exp\left( \int_{boundary} d^d x \phi_0 O \right) \right\rangle = e^{-S[\phi]}, \tag{6}$$

where the left hand side (which is the generating functional $Z$ for the correlation functions in the field theory) is calculated in the boundary, and the exponent on the right-hand side is the action evaluated using the classical solution to the equation of motion for $\phi$ in the bulk with the boundary condition $\phi|_{r=0} = \phi_0$.

There is a $\phi$ for every operator $O$ in the dual field theory. Some examples of the correspondence (that are determined by symmetry) between bulk fields and boundary operators are: The stress-energy tensor $T_{\mu\nu}$ is the response of a local QFT to local change in the metric ($g_{\mu\nu} \leftrightarrow T_{\mu\nu}$). Electromagnetic (gauge) fields in the bulk correspond to currents in the boundary theory ($A_\mu^a \leftrightarrow J_a^\omega$).

As a simple example let us consider a scalar field in the bulk, with action

$$S = -\frac{1}{2}\int d^{d+1}x\sqrt{g}(g^{AB}\partial_A\phi\partial_B\phi + m^2\phi^2), \tag{7}$$

where $\sqrt{g} = \sqrt{|\det g|} = (L/r)^{d+1}$. Note that in our case of a diagonal metric, $g^{AA} = 1/g_{AA}$. We remark that masses in the bulk do not correspond to masses in the dual field theory. The field equation for $\phi$ is given by

$$\frac{1}{\sqrt{g}}\partial_A(\sqrt{g}g^{AB}\partial_B\phi) - m^2\phi = 0, \tag{8}$$

which leads to

$$\left[r^{d+1}\partial_r\left(\frac{1}{r^{d-1}}\partial_r\right) + r^2(-\partial_t^2 + \nabla^2) - m^2L^2\right]\phi = 0. \tag{9}$$

Integrating by parts we can write the action (7) as

$$S = -\frac{1}{2}\int d^{d+1}x[\partial_A(\sqrt{g}g^{AB}\phi\partial_B\phi) - \phi\partial_A(\sqrt{g}g^{AB}\partial_B\phi) + \sqrt{g}m^2\phi^2]. \tag{10}$$

Using Stoke's theorem we find

$$S = -\frac{1}{2}\int_{boundary} d^dx\sqrt{g}g^{rr}\phi\partial_r\phi + \frac{1}{2}\int dx^{d+1}\phi[\partial_A\sqrt{g}g^{AB}\partial_B - m^2]\phi. \tag{11}$$

If $\phi$ solves the equation of motion, the on shell action is given just by the boundary term. Translation invariance allows us to Fourier transform in the directions other than *r*. We write $\phi(\vec{x},r,t) = e^{i(-\omega t+\vec{\mathrm{k}}.\vec{x})}f(r)$ and obtain

$$[r^{d+1}\partial_r(r^{-d+1}\partial_r) - r^2(\omega^2 - \mathrm{k}^2) - m^2L^2]f(r) = 0 \quad . \tag{12}$$

We first consider asymptotic solutions of this equation near the boundary *r* = 0, writing $f(r) \propto r^\Delta$. Taking this expression into Eq. (9) we obtain

$$\Delta(\Delta - d)r^\Delta + r^{2+\Delta}(\omega^2 - \mathrm{k}^2) - m^2L^2r^\Delta = 0. \tag{13}$$

Which for $r \to 0$ leads to

$$\Delta(\Delta - d) = m^2 L^2, \tag{14}$$

with solutions

$$\Delta_{\pm} = \frac{d}{2} \pm \sqrt{\left(\frac{2}{2}\right)^2 + m^2 L^2}. \tag{15}$$

We see that $\Delta_+ \equiv \Delta$ is always positive, therefore $r^{\Delta}$ decays near the boundary and the leading behavior near the boundary of any general solution is $\phi \propto r^{\Delta}$.

Noting that Eq.(12) can be written as

$$\left[ r^2 \frac{d^2}{dr^2} + (1-d) r \frac{d}{dr} - r^2 k^2 - m^2 L^2 \right] f_k(r) = 0, \tag{16}$$

and remembering that the Bessel equation is

$$x^2 \frac{d^2 J_n(x)}{dx^2} + x \frac{dJ_n(x)}{dx} + (x^2 - n) J_n(x) = 0, \tag{17}$$

we see that writing $k^2 = \omega^2 - \mathrm{k}^2$, the solution of Eq. (16) for $k^2 > 0$ is given by

$$f_k(r) = A r^{d/2} K_\nu(kr) + B r^{d/2} I_\nu(kr), \tag{18}$$

where $K_\nu$ and $I_\nu$ are modified Bessel functions, and $\nu = \Delta - d/2 = \sqrt{(d/2)^2 + m^2 L^2}$.

We choose $K$ over $I$ because it is well behaved near the horizon, $K(x) \approx e^{-x}$ at large $x$. As $r \to 0$, $K_\nu(r) \propto r^{-\nu}$. Therefore near the boundary, the bulk solution will behave like:

$$\phi = \phi_{(0)} r^{d-\Delta}, \qquad \text{as } r \to 0. \tag{19}$$

If $d - \Delta \geq 0$ the bulk field $\phi$ goes to a constant or zero at the boundary $r \to 0$.

Choosing a cut off at $r = \varepsilon$ near $r = 0$ (to avoid divergences) we can normalize the solution (18) using the condition $f_k(r = \varepsilon) = 1$, and write

$$f_k(r) = \frac{r^{d/2} K_\nu(kr)}{\varepsilon^{d/2} K_\nu(k\varepsilon)}, \tag{20}$$

which is the ´bulk-to- boundary propagator`. The general solution to Eq.(9) can be written as

$$\phi(r,x)=\int\frac{d^dk}{(2\pi)^d}e^{ik.x}f_k(r)\phi_0(k), \tag{21}$$

where $\phi_0(k)$ is determined by the boundary condition

$$\phi(\varepsilon,x)=\int\frac{d^dk}{(2\pi)^d}e^{ik.x}\phi_0(k). \tag{22}$$

The field $\phi$ is a classical solution that is regular in the bulk and asymptotes to a given value $\phi_0$ at the boundary. The action on the shell reduces to the surface terms

$$S=\frac{1}{2}\int\frac{d^dk}{(2\pi)^d}\phi_0(-k)\Im(k,r)\phi_0(k)\Big|_{r=\varepsilon}^{r=\infty}, \tag{23}$$

where

$$\Im(k,r)=\sqrt{g}g^{rr}f_k(r)\partial_r f_k(r). \tag{24}$$

The two point correlation function evaluated using the equation

$$<O(x)O(y)>=\frac{\delta}{\delta J(x)}\frac{\delta}{\delta J(y)}\ln Z\Big|_{J=0}, \tag{25}$$

becomes

$$<O(k_1)O(k_2)>=-\frac{\delta}{\delta\phi_0(k_1)}\frac{\delta}{\delta\phi_0(k_2)}S=(2\pi)^d\delta^d(k_1+k_2)\Im(k_1,\varepsilon). \tag{26}$$

The calculation is presented in Ref.[18] and the final result is

$$<OO>_{\mathrm{k},\omega}\propto(\mathrm{k}^2+\omega^2)^{\Delta_+-d/2}. \tag{27}$$

It is know that if for large $x$ we have

$$<O(x)O(0)>=\frac{1}{|x|^{2h}}, \tag{28}$$

the parameter $h$ is called the scaling dimension of the operator $O$. Fourier transforming (28) we find

$$<OO>\propto k^{2(h-d/2)}. \tag{29}$$

Comparing with Eq. (27) we see that $\Delta$ can be interpreted as the scaling dimension of the operator $O$ in the boundary theory.

If $\Delta \leq d$ (i.e. $d - \Delta \geq 0$) the operator $O$ is relevant or marginal. Relevant operators do not destroy the asymptotically AdS region of the metric [16].

**4. Finite temperatures**

Finite temperature or chemical potential breaks the dilatation symmetry of the space time. However, we expect that the space time should recover scaling invariance as we go towards the boundary, this is, the space time should be asymptotically Anti-de Sitter. If we relax the scaling symmetry but wish to preserve spatial rotations and space time translations, we start with the metric written as

$$ds^2 = \frac{L^2}{r^2}\left(-f(r)dt^2 + g(r)dr^2 + h(r)dx^i dx^i\right). \tag{30}$$

If $f \neq h$ the metric breaks Lorentz invariance, as is the case for finite temperature or finite chemical potential. We know that all quantum field theory can be placed at temperatures different from zero. Therefore we do not need a new ingredient here and can use the same action as the one used in the $T = 0$ case. Taking the metric (30 ) in the Einstein equation of motion (5) one finds:

$$ds^2 = \frac{L^2}{r^2}\left(-f(r)dt^2 + \frac{dr^2}{f(r)} + dx^i dx^i\right), \tag{31}$$

with

$$f(r) = 1 - \left(\frac{r}{r_+}\right)^d, \tag{32}$$

where $r_+$ is a constant, which we will interpret below. Since $f \to 1$ as $r \to 0$, this space time is asymptotically AdS.

Before going ahead, I will present a small comment. Let us consider a metric written as $ds^2 = -g_{tt}dt^2 + g_{rr}dr^2$. The light cones are given by the condition $ds^2 = 0$, which leads to

$$\frac{dt}{dr} = \pm\sqrt{\frac{g_{rr}}{g_{tt}}}. \tag{33}$$

If $dt/dr \to \pm\infty$ the light cones close up. Thus a light ray that approaches the singularity never seems to get there, at least in the above coordinate system. So there is a horizon at $r = r_+$. The surface at $r = r_+$ is infinitely redshifted with respect to an asymptotic observer. This solution is generally called by the name of ´black hole`, although in our case it should be better called a ´black membrane`. Events at $r > r_+$ cannot influence the boundary near $r = 0$.

As it is well known the partition function at finite temperature $T$ is expressed as a Euclidean path integral over periodic Euclidean time path. Thus the finite temperature field theory is obtained by considering periodic imaginary time. We can use this approach to deduce that black holes radiate thermally at a given temperature $T$.

To proceed we perform an analytic continuation to Euclidian time $\tau = it$ and write the metric as

$$ds_E^2 = \frac{L^2}{r^2}\left( f(r)d\tau^2 + \frac{dr^2}{f(r)} + dx^i dx^i \right). \tag{34}$$

Near $r = r_+$ we can write

$$f(r) \approx f(r_+) + (r - r_+) f`(r_+) = (r - r_+) f`(r_+), \tag{35}$$

and the metric takes the following form in the vicinity of the horizon

$$ds_E^2 \approx \frac{f`(r_+)(r - r_+)}{r_+^2} d\tau^2 + \frac{dr^2}{r_+^2 f`(r_+)(r - r_+)} + \frac{dx^i dx^i}{r_+^2}. \tag{36}$$

This metric has a coordinate singularity at the horizon, which we can remove by a suitable choice of coordinates [23]. This is, at $r = r_+$ the Euclidean time direction shrinks to a point. This is similar to what happens at the origin of polar coordinates

$$ds_2 = d\rho^2 + \rho^2 d\varphi^2. \tag{37}$$

At ρ = 0 the metric is singular, but we know that the geometry is regular provided φ has period 2π. If $\varphi \in [0, 2\pi - \delta]$ with $\delta \neq 0$, ρ = 0 is a singular point, and the metric describes a cone. Changing coordinates so that the $\{r,t\}$ part of the Euclidean metric look like the $\{\rho,\varphi\}$ metric near the horizon we find

$$d\rho^2 = \frac{dr^2}{r_+^2 f`(r_+)(r - r_+)}, \tag{38}$$

$$\rho = \frac{2}{r_+\sqrt{f`(r_+)}}. \tag{39}$$

Writing $\varphi = \beta\tau$ we find $\beta = f`(r_+)/2$ where $|f`(r_+)| = d/r_+$.

This metric describes a regular geometry if $\varphi$ is periodic with period 2π:

$$\beta\tau \to \beta\tau + 2\pi \quad \tau \to \tau + \frac{2\pi}{\beta} = \tau + \frac{4\pi r_+}{d}. \tag{40}$$

Identifying the temperature as the inverse of the periodicity we get

$$T = \frac{d}{4\pi r_+}. \tag{41}$$

So a QFT in the presence of a black hole has a temperature $T$. Generally, the specific heat of a black hole is negative and the black hole is thermodynamically unstable, but in the case of a AdS space the black hole is stable and therefore represents an equilibrium situation. Notice that at zero temperature $r_+ = \infty$, and we recover the AdS metric. The area of the horizon (the set of points with $r = r_+$ and $t$ fixed) is

$$A = \int_\Sigma \sqrt{g} d^{d-1}x = \left(\frac{L}{r_+}\right)^{d-1} V \quad , \tag{42}$$

where $V$ is the spatial volume of the boundary $\Sigma_{d-1}$ ($\Sigma_{d-1}$ is some manifold with dimension d-1. We can give it finite volume as an IR regulator). The entropy is given by

$$\mathrm{S} = \frac{A}{4G_N} = \frac{L^{d-1}V}{4G_N r_+^{d-1}} = \frac{L^{d-1}V}{4G_N}\left(\frac{4\pi T}{d}\right)^{d-1} . \tag{43}$$

The entropy density $s = \mathrm{S}/V$ is

$$s = \frac{a}{4G_N}, \tag{44}$$

where $a \equiv A/V$ is the area per unit volume (area density). The calculation of the entropy for a strongly interacting QFT is generally difficult. Using the AdS/CFT correspondence the entropy can be found simply through calculating the area of the horizon.

Dissipation due to the finite temperature is described by matter falling through the black hole horizon.

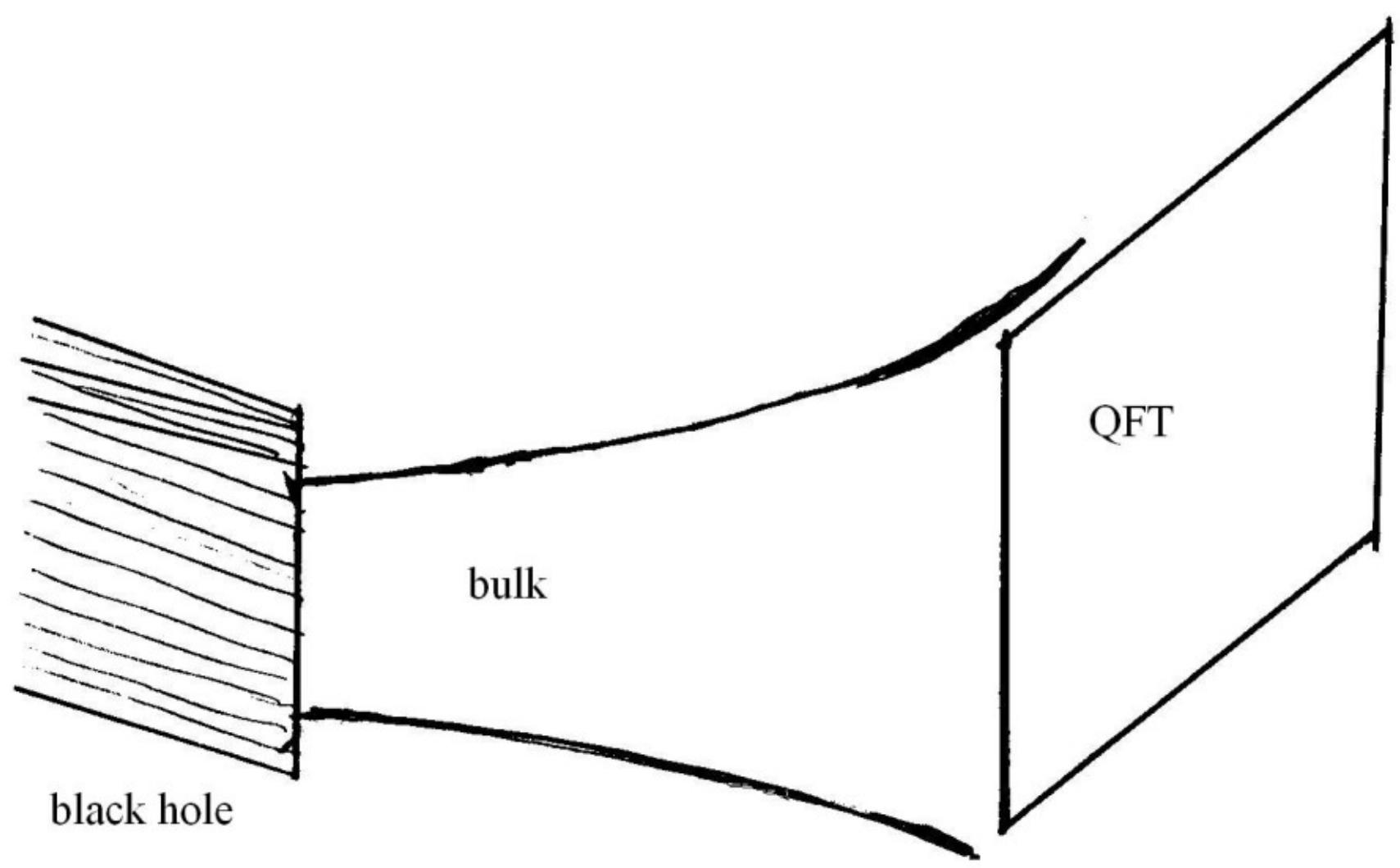

Figure 2: The QFT at finite temperature is dual to classical gravity in a spacetime with a black hole.

Now I will calculate the entropy from the action. The on-shell gravity action for the black hole solution is given by

$$S = S_E + S_{GB} + S_{ct}, \tag{45}$$

where

$$S_E = -\frac{1}{16\pi G_N}\int d^{d+1}x\sqrt{g}\left[R + \frac{d(d+1)}{L^2}\right], \tag{46}$$

is the bulk gravity Euclidean action (see Eq. A11), the second term is the Gibbon-Hawking boundary term (see Eq. A12), and $S_{ct}$ is an intrinsic boundary counter-term, needed to subtract divergences as $r \to 0$, and render the action finite, given by [16]:

$$S_{ct} = \frac{1}{8\pi G_N}\int_{\partial\Sigma} d^d x\sqrt{\gamma}\,\frac{(d-1)}{L}. \tag{47}$$

The metric $\gamma$ is defined by

$$\underset{r\to 0}{ds^2} \approx \frac{L^2}{r^2}dr^2 + \gamma_{\mu\nu}dx^\mu dx^\nu. \tag{48}$$

Using $R = -(d+1)d/L^2$, we can write

$$S_E = -\frac{1}{16\pi G_N}\frac{d}{L^2}\int d^{d+1}x\sqrt{g}. \tag{49}$$

We have seen that $\sqrt{g} = (L/r)^{d+1}$. The integral in $r$ is calculated from a cut off $r = \varepsilon$, up to $r = r_+$ and after that we integrate on a space with geometry $S^1 \times \Sigma_{d-1}$, where $S^1$ has radius 1 / $T$, and $\Sigma_{d-1}$ is a manifold with dimension $d$ -1. The other terms are easily calculated at $r = \varepsilon$. The final result is

$$S = -\frac{L^{d-1}}{16\pi G_N}\frac{V}{r_+^d T}, \tag{50}$$

where $V$ is the volume of $\Sigma_{d-1}$. From Eq. (50) we obtain the free energy $F = TS$ and the entropy $S = -\partial F/\partial T$. We find the same result given by Eq. (43).

## 5. Finite chemical potential

The standard action for the Maxwell theory is given by

$$S_M = -\frac{1}{4g^2}\int d^{d+1}x\sqrt{g}F_{\mu\nu}F^{\mu\nu}, \tag{51}$$

where $F_{\mu\nu} \equiv \partial_\mu A_\nu - \partial_\nu A_\mu$, and $g$ is the Maxwell coupling. The Maxwell equation is

$$\nabla_\mu F^{\mu\nu} = 0, \tag{52}$$

and the Einstein equation of motion are now given by

$$R_{\mu\nu} - \frac{R}{2}g_{\mu\nu} - \frac{d(d-1)}{2L^2}g_{\mu\nu} = \frac{k^2}{2g^2}\left(2F_{\mu\eta}F_\nu^\eta - \frac{1}{2}g_{\mu\nu}F_{\alpha\beta}F^{\alpha\beta}\right), \tag{53}$$

where $k^2 = 8\pi G_N$. We will consider a Maxwell field where we have only the scalar potential $A_t(r)$. Taking the metric (31) in the above equations we find

$$f(r) = 1 - \left(1 + \frac{r_+^2\mu^2}{\gamma^2}\right)\left(\frac{r}{r_+}\right)^d + \frac{r_+^2\mu^2}{\gamma^2}\left(\frac{r}{r_+}\right)^{2(d-1)}, \tag{54}$$

where

$$\gamma^2 = \frac{(d-1)g^2L^2}{(d-2)k^2}, \tag{55}$$

and

$$A_t = \mu\left[1 - \left(\frac{r}{r_+}\right)^{d-2}\right]. \tag{56}$$

The horizon radius is given by the largest real root of $f(r) = 0$. The above solution is known as the Reissner-Nordstrom-AdS black hole, and is dual to a $d$ – dimensional strongly-coupled field theory at finite temperature $T$ and finite charge density. The asymptotic value of the bulk gauge field $A_t(0) = \mu$ is interpreted in the dual theory as the chemical potential for the electric charge density.

The chemical potential μ is the energy needed to add one particle to a thermally and mechanically isolated system. For a particle with charge $e$ in an electric potential $A_t$ we have $\mu = eA_t$. We remark that as $A_t$ depends only on $r$, the electric field is in the $r$ direction in the bulk and the rotational symmetry is preserved.

The temperature can be calculated using the analytic continuation to a Euclidean metric, and we obtain

$$T = \frac{1}{4\pi r_+}\left[d - \frac{(d-2)r_+^2\mu^2}{\gamma^2}\right]. \tag{57}$$

A background magnetic field $B = F_{(0)xy}$ preserves rotational symmetry only in a 2+1 dimension field theory. Adding a magnetic field in the case $d$ =3 is straightforward. One finds a dyonic black hole, i. e. a black hole with both electric and magnetic charge [16].

**6. Linear response and transport coefficients**

Here we will study the response of a system where we have an external field $\phi_0(x)$ coupled to an operator $O(x)$, with the following interaction Hamiltonian

$$H_{\text{int}} = -\int d^d x \phi_0(t,\vec{x}) O(t,\vec{x}) \ . \tag{58}$$

For simplicity we perturb with the operator $O$, and measures the operator $O$. As it is well known from the linear response theory, $\phi_0$ produces a change in the expectation value of $O$ given by

$$\delta < O >= \int d^{d+1} x` G^R(x,x`)\phi_0(x`), \tag{59}$$

where

$$G^R(x,x`) = i\theta(t-t`) < [O(x), O(x')] >, \tag{60}$$

is the retarded Green`s function. Fourier transforming we find

$$< O(\omega,\vec{k}) >= -G^R(\omega,\vec{k})\phi(\omega,\vec{k}). \tag{61}$$

We define a transport coefficient $\chi$ by

$$\chi = -\lim_{\omega \to 0} \lim_{\vec{k} \to 0} \frac{1}{\omega} G^R(\omega,\vec{k}). \tag{62}$$

In the case where $O = j^\mu$ is a conserved current, $\phi_0 = A_\mu$ is the boundary behavior of a bulk gauge field, and the transport coefficient is the conductivity

$$\delta < \vec{J} > \underset{k \to 0, \omega \to 0}{\to} i\omega\chi(j)\vec{A} = \sigma\vec{E}. \tag{63}$$

In the classical mechanics the derivative of an on-shell action with respect to the boundary value of a field is simply equal to the canonical momentum conjugate to the field, evaluated at the boundary. This suggest the following definition

$$< O >= \frac{\delta S[\phi_0]}{\delta \phi_0} \equiv \lim_{r \to 0} \Pi[\phi_0]. \tag{64}$$

To obtain a finite action we need to add boundary terms to the action ( see[16] and appendix A).We can think of $\Pi$ as the bulk field-momentum, with $r$ thought as time.

$$G^R_{OO} = \lim_{r \to 0} \frac{\delta \Pi}{\delta \phi}\bigg|_{\delta\phi=0} . \tag{65}$$

**7. Calculation of transport coefficients**

Here I will discuss the bulk calculation of transport coefficients following the discussion of [16] and [17]. To obtain the retarded Green´s function at finite temperature, one solves the bulk equations of motion of the field $\phi$ dual to $O$, and linearize the equations about the black hole metric. Iqbal and Liu [17] consider a fictitious membrane at each constant-radius hypersurface (starting at the black hole) and introduce a linear response function for each of them. They then derive a flow equation for the radius-dependent response function, and calculate how it evolves from the horizon to the boundary, where it determines the response of the dual field theory. To do that they start with the metric (31) written in a general form as

$$ds^2 = -g_{tt}dt^2 + g_{rr}dr^2 + g_{ij}dx^i dx^j \ . \tag{66}$$

Near the horizon $r = r_+$ the metric may be written

$$g_{tt} = A(r - r_+), \qquad\qquad g_{rr} = \frac{B}{r - r_+}. \tag{67}$$

For our case $A = L^2 d / r_+^3,\ B = L^2 / r_+ d$, but for the moment I will keep the general result (66). The equation of motion (9), with $m = 0$, for this metric is given by

$$\frac{\partial}{\partial r}\left(\frac{r - r_+}{B}\frac{\partial \phi}{\partial r}\right) - \frac{1}{A(r - r_+)}\frac{\partial^2 \phi}{\partial t_2} = 0. \tag{68}$$

Taking $\phi(r,t) = \phi(r)e^{i\omega t}$ leads to

$$\frac{A}{B}(r - r_+)\frac{\partial}{\partial r}(r - r_+)\frac{\partial \phi(r)}{\partial r} + \omega^2 \phi(r) = 0. \tag{69}$$

Introducing a variable $x$ such that

$$\frac{dx}{dr} = \sqrt{\frac{g_{rr}}{g_{tt}}}, \tag{70}$$

and using

$$\frac{g_{rr}}{g_{tt}} = \frac{B}{(r-r_+)^2 A}, \tag{71}$$

we find

$$\frac{\partial^2 \phi(x)}{\partial x^2} + \omega^2 \phi(x) = 0. \tag{72}$$

The solution is $\phi(x,t) \propto e^{-i\omega(t \pm x)}$.

The in-falling boundary condition implies that we should take the positive sign in the exponent. Using a coordinate $v$ defined by $v = x + t$, and integrating the equation

$$dv = dt + \sqrt{\frac{g_{rr}}{g_{tt}}} dr, \tag{73}$$

we obtain

$$v = t + \frac{1}{4\pi T} \ln(r - r_+), \tag{74}$$

where I have used

$$\frac{1}{T} = 4\pi \sqrt{\frac{A}{B}}. \tag{75}$$

The final result is

$$\phi(r,t) \propto (r - r_+) \exp\left(-\frac{i\omega}{4\pi T}\right) e^{-i\omega t}. \tag{76}$$

In the coordinates ($v$, $r$) the metric is written as

$$ds^2 = A(r - r_+) dv^2 - 2\sqrt{AB}\, dv dr. \tag{77}$$

In these coordinates the metric is non singular at $r = r_+$.

To proceed we start with the action at the horizon

$$S_H = -\int_\Sigma d^d x \frac{\sqrt{g}}{q(r)} g^{rr} (\partial_r \phi) \phi(r_+, x), \tag{78}$$

where $q(r)$ is an effective scalar coupling. In practice, Σ is a time like surface of fixed $r$ just outside the true horizon. Multiplying and dividing by $\sqrt{\gamma}$ where $\gamma_{\mu\nu}$ is the induced metric on the horizon Σ we have

$$S_H = \int_\Sigma d^d x\sqrt{\gamma}\left[-\frac{\sqrt{g}}{\sqrt{\gamma}q(r)}g^{rr}\partial_r\phi\right]\phi(r_+,x)\,. \tag{79}$$

Using Eq.(64) we see that the momentum conjugate to $\phi$ ($r$ direction) is

$$\Pi = -\frac{\sqrt{g}}{q(r)}g_{rr}\partial_r\phi\,. \tag{80}$$

The horizon is a regular place for in-falling observers. For these observers $\phi$ is not singular. This means that $\phi$ can depend on $r$ and $t$ only through the coordinate $v$, this is $\phi(r,t,x_i) = \phi(v,x_i)$. We have

$$\frac{d}{dr} \to \sqrt{\frac{g_{rr}}{g_{tt}}}\partial_t\phi, \tag{81}$$

leading to

$$\Pi = -\frac{\sqrt{g}}{q(r)}g^{rr}\partial_t\phi. \tag{82}$$

Using $g^{rr} = 1/g_{rr}$, and $\phi \propto e^{-i\omega t}$, we arrive at the final result

$$\Pi(r,k_\mu) = \frac{1}{q(r)}\sqrt{\frac{g}{g_{rr}g_{tt}}}i\omega\phi(r_+,k_\mu). \tag{83}$$

It can be shown [17] that in the limit $k_\mu \to 0,\ \omega \to 0,$ we have

$$\left.\frac{\Pi}{\omega\phi}\right|_{r=0} = \left.\frac{\Pi}{\omega\phi}\right|_{r=r_+}. \tag{84}$$

Thus

$$\chi = \frac{1}{q(r_+)}\left.\sqrt{\frac{g}{g_{rr}g_{tt}}}\right|_{r_+}, \tag{85}$$

where

$$g = g_{tt}g_{rr}\left(\frac{1}{r_+^2}\right)^{d-1}. \tag{86}$$

In the case where $O = T_y^x$, the source is the boundary value of a metric perturbation $\delta g_x^y$, and the transport coefficient χ is the shear viscosity η. Viscosity is associated with the tendency of a substance to resist flow. A perfect fluid has negligible shear viscosity. Lower shear viscosities are associated with strongly interacting particles. As we have seen, the entropy density $s$ of the boundary field theory is given by $s = A/4G_N V_{d-1}$, which leads to

$$\frac{\eta}{s} = \frac{4G_N}{q(r_+)}. \tag{87}$$

In the case of Einstein gravity $q = 16\pi G_N$, and we get

$$\frac{\eta}{s} = \frac{1}{4\pi}. \tag{88}$$

The ratio η /*s* gives a measure of the interaction per constituent that better allows comparison across different systems at widely different scales. The horizon response always corresponds to that of the low frequency limit of the boundary theory (regardless of the specific model we use). Away from this limit, the full geometry of the space-time becomes important.

Iqbal and Liu [17] have also derived expressions for various transport coefficients in terms of components of the metric evaluated at the horizon. These coefficients are determined by universal constants of nature, and not by collision rate.

## 8. Dynamics close to equilibrium

In this section I will study the heat current in the context of the AdS/CFT correspondence. For simplicity I will take $d$ = 3, zero chemical potential and consider the limit of zero momentum. The heat current (which is the response of the system to a temperature gradient) can be written as

$$< Q_x > = < T_{tx} > = -k\frac{\partial T}{\partial x}, \tag{89}$$

where $k$ is the thermal conductivity and I have taken the current pointing in the $x$ direction.

The period of Euclidian time is 1 / $T$, and therefore we can rescale the time so that there is no temperature dependence in the period. Writing $t$ = t /$T$, the metric, in the Minkowski space and in this coordinate, becomes $g_{\mathrm{tt}} = -1/T^2$. A small thermal gradient $T \to T + x\partial_x T$ implies

$$\delta g_{\mathrm{tt}(0)} = -\frac{2x\partial_x T}{T^3}. \tag{90}$$

Going back to the original time $t$, we have

$$\frac{\partial g_{tt(0)}}{\partial x} \approx -\frac{2\partial_x T}{T}. \tag{91}$$

Now I take $g_{tt}$ constant and perform a change of coordinates in which the temperature gradient is exhibited by a fluctuation of an off-diagonal component of the metric (the energy transport is related to the $T_{tx}$ component of the stress-energy tensor). Under an infinitesimal coordinate transformation $x^\alpha \to x^\alpha + \xi^\alpha$, the metric changes by

$$\delta g_{\mu\nu} = \partial_\mu \xi_\nu + \partial_\nu \xi_\mu . \tag{92}$$

This can be verified using the transformation law for the metric components

$$g_{\hat{\mu}\hat{\nu}} = \frac{\partial x^\mu}{\partial x^{\hat{\mu}}} \frac{\partial x^\nu}{\partial x^{\hat{\nu}}} g_{\mu\nu} . \tag{93}$$

Following Herzog [18] we can choose $\xi_\mu$ such that $\partial_x (g_{tt} + \delta g_{tt}) = 0$. Setting $\xi_x = 0$, and using $\delta g_{tx} = \partial_t \xi_x + \partial_x \xi_t = \partial_x \xi_t$, we obtain $\partial_x g_{tt} + \partial_x \delta g_{tt} = 0$. But $\partial_x \delta g_{tt} = \partial_x (\partial_t \xi_t + \partial_t \xi_t) = 2\partial_t \partial_x \xi_t$, leading to $2\partial_t \partial_x \xi_t = -\partial_x g_{tt}$. If we take a time dependence of the form $e^{-i\omega t}$ we obtain

$$\partial_x \xi_t = \frac{\partial_x g_{tt(0)}}{2i\omega} = -\frac{\partial_x T}{i\omega T} \ , \tag{94}$$

which gives

$$\delta g_{tx(0)} = -\frac{\partial_x T}{i\omega T} . \tag{95}$$

From Eqs. (89) and (95) we can write

$$< Q_x >= kTi\omega \delta g_{tx(0)} \ . \tag{96}$$

The next step is to solve the Einstein's equation of motion of perturbation $\delta g_{tx}$ in the bulk. Linearising this equation about the background solution of the four dimensional black hole, we find

$$\frac{dh}{dr} + \frac{2h}{r} = 0 , \tag{97}$$

where $h = \delta g_{tx}$. The solution is $h = ar^{-2}$, where $a$ is a constant. Using the action (45) we obtain

$$\Pi_{g_{tx}} = \frac{\delta S}{\delta g_{tx(0)}} = \frac{L^2}{4\pi G_N r^3} \left( 1 - \frac{1}{\sqrt{f}} \right) \delta g_{tx(0)} \ , \tag{98}$$

where $f$ is given by Eq. (32), and $g_{tx(0)}$ is defined at finite $r$ by $g_{tx(0)} = (r/L)^2 g_{tx}$. Using Eq.(64) we obtain in the limit $r \to 0$

$$< T_{tx} >= -\varepsilon \delta g_{tx(0)} \ , \tag{99}$$

where $\varepsilon = L^2/8\pi G_N r_+^3$ . Comparing (96 ) and (99) gives

$$k(\omega) = \frac{i\varepsilon}{\omega T} \ . \tag{100}$$

The case of a finite chemical potential was studied by Hartnoll [16].

Recently, Brynjolfsson et al. [24] have shown that gravitational dual models with dynamical critical exponent $z > 1$ can produce a non Fermi liquid behavior for the specific heat of a quantum critical system that is qualitatively similar to what is seen in experiments on heavy fermions systems.

**9. Conclusions**

In this conclusion I will discuss briefly spin transport in the model presented in Sec.2. More details can be found in Refs.[25] and [19]. Spin transport is a very interesting problem in condensed matter physics. In a insulating magnet, magnetization can be transported by excitations such as magnons and excitons without transport of charge. As it is expected that transport properties in the critical region should have a universal character, we can use a CFT, which is solvable by the AdS/CFT correspondence, to get results that can be used in our model.

While charge conductivity is studied as the current response to a time-dependent electromagnetic potential, the spin current flows in response to a magnetic field gradient, which plays a role of the chemical potential for spins. Spin transport for the model described by the Hamiltonian (1) was studied for $D > D_C$ and $D < D_C$ in Ref.[25] and [26] respectively. In the critical region, spin transport can be studied using an expression similar to Eq.(2), including the effect of the magnetic field gradient [19].

The spin conductivity, $\sigma(\omega)$, is related to the retarded correlation function, $\chi(k,\omega)$, by the Kubo formula

$$\sigma(\omega) = -\lim_{k\to 0} \frac{i\omega}{k^2}\chi(k,\omega). \tag{101}$$

It can be shown that at high frequencies or low temperatures ( $\omega / T >> 1$), $\chi(k,\omega)$, in all CFT in spatial dimension higher than one, reduces to the $T = 0$ limit, where scale invariance, Lorentz symmetry and "charge" conservation determines $\chi(k,\omega)$ up to a constant [5]:

$$\chi(k,\omega) = const.\frac{Kk^2}{\sqrt{\mathrm{v}^2k^2 + (\omega + i\eta)^2}}, \qquad \hbar\omega >> k_B T, \tag{102}$$

where $K$ is a universal number, and v is the speed of the excitations.

In the low frequency limit ( $\hbar\omega << k_B T$) one expects a diffusive behavior, and we have

$$\chi(k,\omega)=Q^2\,\frac{\chi_C Dk^2}{Dk^2-i\omega}, \qquad (103)$$

where $\chi_C$ is the compressibility, $D$ is the diffusion constant and for the spin transport the "charge" $Q$ is given by [25] $Q=g\mu_B/4\sqrt{\hbar}$. Quantum critical scaling arguments show that [5]

$$\chi_C=\Theta_1 k_B T/(h\text{v})^2, \qquad D=\Theta_2 h\text{v}^2/(k_B T), \qquad (104)$$

where $\Theta_1$ and $\Theta_2$ are universal numbers.

The first exact results for a quantum- critical point in 2+1 dimensions was obtained in a supersymmetric Yang-Mills non-Abelian gauge theory, using the AdS/CFT correspondence [27]. Exact results for the full structure of $\chi(k,\omega)$ was obtained and the expected limiting forms in Eqs. (102) and (103) was obeyed. Exact results for $K$, $\Theta_1$ and $\Theta_2$ were found. A curious feature in this theory was the relation $K=\Theta_1\Theta_2$.

**Appendix A: Basics of general relativity**

We start with the line element given by

$$ds^2=g_{\mu\nu}dx^\mu dx^\nu, \qquad (A.1)$$

where $g_{\mu\nu}$ is the metric tensor. It is common to use the term *metric* to $ds^2$. We define the inverse metric $g_{\mu\nu}$, via $g^{\mu\nu}g_{\nu\sigma}=\delta^\nu_\sigma$. The metric can be used to lower and rise indices: $x^\mu=g^{\mu\nu}x_\nu$, $x_\mu=g_{\mu\nu}x^\nu$.

Using the metric tensor we define an object called the Christoffel symbols by:

$$\Gamma^\lambda_{\mu\nu}=\frac{1}{2}g^{\lambda\sigma}(\partial_\mu g_{\nu\sigma}+\partial_\nu g_{\sigma\mu}-\partial_\sigma g_{\mu\nu}). \qquad (A.2)$$

The next object is the Riemann tensor

$$R^\rho_{\sigma\mu\nu}=\partial_\mu\Gamma^\rho_{\nu\sigma}-\partial_\nu\Gamma^\rho_{\mu\sigma}+\Gamma^\rho_{\mu\lambda}\Gamma^\lambda_{\nu\sigma}-\Gamma^\rho_{\nu\lambda}\Gamma^\lambda_{\mu\nu}, \qquad (A.3)$$

and

$$R_{\rho\sigma\mu\nu}=g_{\rho\lambda}R^\lambda_{\sigma\mu\nu}. \qquad (A.4)$$

The Riemann tensor can be used to derive two quantities. The first is the Ricci tensor, which is calculated by contraction on the first and third indices:

$$R_{\mu\nu}=R^\lambda_{\mu\lambda\nu}. \qquad (A.5)$$

Using contraction on the Ricci tensor, we obtain the Ricci scalar

$$R = R^{\mu}_{\mu} = g^{\mu\nu} R_{\mu\nu}. \tag{A.6}$$

The Einstein`s equations for the gravity are then written as

$$R_{\mu\nu} - \frac{1}{2} R g_{\mu\nu} + \Lambda g_{\mu\nu} = 8\pi G T_{\mu\nu}, \tag{A.7}$$

where $\Lambda$ is the cosmological constant and $T_{\mu\nu}$, the *stress-energy* or *energy-momentum* tensor, acts as the source of the gravitational field. The $T^{00}$ component represents energy density. The $T^{i0}$ component is the momentum density in the *i* direction. The component $T^{0i}$ (which is equal to $T^{i0}$) represents the energy flux across the surface $x^{i}$. And finally $T^{ij}$ is the *stress*.

The calculation of the Christoffel symbols, the Riemann tensor and other objects, starting from the metric, can be performed using programs such as *Maple* and *Mathematica*.

In the empty space (i.e. $T_{\mu\nu} = 0$) Eq. (A.7) becomes

$$R_{\mu\nu} - \frac{1}{2} R g_{\mu\nu} + \Lambda g_{\mu\nu} = 0. \tag{A.8}$$

Multiplying (A.8) by $g^{\mu\nu}$, using Eq.(A.6) and $g^{\mu\nu} g_{\mu\nu} = n$, where *n* is the dimension of the space time, we find

$$R = \frac{2n}{n-2} \Lambda. \tag{A.9}$$

Taking (A.9) into (A.8) we obtain

$$R_{\mu\nu} = \frac{2\Lambda}{n-2} g_{\mu\nu}. \tag{A.10}$$

The Einstein´s equation can be derived from the following action

$$S = \int d^4x \sqrt{g} \left[ \frac{1}{16\pi G_N} (R - 2\Lambda) \right]. \tag{A.11}$$

For an introduction to general relativity see [28]. When the underlying space time has a boundary one needs to add a term called Gibbons-Hawking boundary term [29], so that the variational principle is well-defined. This term affects which boundary conditions we impose on the metric, in the same way that the $\int \phi \vec{n}.\vec{\partial} \phi$ term in the scalar case changes the scalar boundary condition from Newmann to Dirichlet. The term is written as

$$S_{GHY} = -\frac{1}{8\pi G_N}\int_{\partial M} d^3x\sqrt{\gamma}\mathrm{K}, \qquad \text{(A.12)}$$

where γ is induced metric on the boundary surface, K is trace of the extrinsic curvature of the boundary

$$\mathrm{K} = \gamma^{\mu\nu}\nabla_\mu n_\nu = \frac{n^r}{2}\gamma^{\mu\nu}\partial_r\gamma_{\mu\nu}, \qquad \text{(A.13)}$$

and $n^r$ is the outward-pointing unit normal to the boundary.

**Appendix B: Anti de Sitter space.**

Anti- de Sitter space (AdS) belongs to a wide class of homogeneous spaces that can be defined as quadric surfaces in flat vector spaces. The standard example is the *d* dimensional sphere $S^d$ defined by

$$X_1^2 + \ldots + X_{d+1}^2 = L^2, \qquad \text{(B.1)}$$

embedded in an Euclidian $d + 1$ dimensional space. The *d*- dimensional de Sitter space is defined by:

$$-X_0^2 + X_{d+1}^2 + \sum_{i=1}^{d-1} X_i^2 = L^2, \qquad \text{(B.2)}$$

embedded in a flat $d + 1$ dimensional space with metric:

$$ds^2 = -dX_0^2 + dX_{d+1}^2 + \sum_{i}^{d-1} dX_i^2. \qquad \text{(B.3)}$$

It has constant positive curvature. The *d* dimensional anti-de Sitter space can be defined as the quadric

$$X_0^2 + X_{d+1}^2 - \sum_{i=1}^{d-1} X_i^2 = L^2, \qquad \text{(B.4)}$$

embedded in a flat $d + 1$ dimensional space with the metric

$$ds^2 = -dX_0^2 - dX_{d+1}^2 + \sum_{i=1}^{d-1} dX_i^2. \qquad \text{(B.5)}$$

It has constant negative curvature. The metric can be written in different forms, corresponding to different coordinate systems. In the Poincaré coordinates $(u, t, \vec{x})$ $(0 < u, \vec{x} \in R^{d-2})$ defined by [30]:

$$X_0 = \frac{1}{2u}[1 + u^2(L^2 + \vec{x}^2 - t^2)], \qquad X_d = Lut,$$

$$X^i = Lux^i, \qquad (i = 1, \ldots, d-2),$$

$$X^{d-1} = \frac{1}{2u}[1 - u^2(L^2 - \vec{x}^2 + t^2)], \tag{B.6}$$

the anti-de Sitter metric becomes

$$ds^2 = L^2\left[\frac{du^2}{u^2} + u^2(-dt^2 + d\vec{x}^2)\right]. \tag{B.7}$$

The Poincaré coordinates cover only part of the AdS space. Another equivalent form of the metric is obtained from (B.7) by setting $r = 1/u$, giving

$$ds^2 = \frac{L^2}{r^2}\left[-dt^2 + dr^2 + d\vec{x}^2\right] \tag{B.8}$$

## Acknowledgments

This work was partially supported by Conselho Nacional de Desenvolvimento Cientifico e Tecnologico (CNPq).

## References

[1] J. M. Maldacena, “The large N limit of superconformal field theories and supergravity,” Adv. Theor. Math. Phys. 2, (1998)231 [arXiv:hep-th/9711200]; I. R. Klebanov and J. M. Maldacena, “Solving quantum field theories via curved spacetimes,” Physics Today, January 2009, p. 28; C. V. Johnson and P. Steiberg, “What black holes teach about strongly coupled particles,” Physics Today, May 2010, p. 29.

[2] S. S. Gubser, I. R. Klebanov and A. M. Polyakov, “Gauge theory correlators from non-critical string theory,” Phys. Lett. B 428(1998) 105 [ArXiv:hepth/980.2109].

[3] E. Witten, “Anti-de Sitter space and holography,” Adv. Theor. Mat. Phys. 2 (1998) 253 [arXiv:hep-th/980.2150].

[4 ] J. McGreevy, “Holographic duality with a view toward many-body physics,” [arXiv:hep-th/0909.05182].

[5] C. P. Herzog, P. Kovtun, S. Sachdev and D. T. Son, “Quantum critical transport, duality, and M-Theory”, Phys. Rev. D75 (2007) 085020 [arXiv:hep-th/0701.036].

[ 6] S. A. Hartnoll and P. Kovtun, “Hall conductivity from dyonic black holes”, Phys. Rev. D76 (2007) 066001 [arXiv:hep-th/0704.1160].

[7] S. A. Hartnoll, P. K. Kovtun, M. Muller and S. Sachdev, “Theory of the Nerst effect near quantum phase transitions in condensed matter, and in dyonic black holes”, Phys. Rev. B76 (2007) 144502 [arXiv:cond-mat/0706.3215].

[8] S. A. Hartnoll and C. P. Herzog, “Ohm`s Law at strong coupling: S duality and cyclotron resonance”, Phys. Rev. D76 (2007) 106012 [arXiv:hep-th/0706.3228].

[9] S. A. Hartnoll and C. P. Herzog, “Impure AdS/CFT”, Phys. Rev. D77 (2008) 106009 [arXiv:hep-th/0801.1693].

[10] T. Faulkner, H. Liu. J. McGreevy, and D. Vegh, “Emergent quantum criticality, Fermi surface, and $AdS_2$”, [arXiv:hep-th/0907.2694].

[11] M. Cubrovic, J. Zanen, and K. Schalm, “String theory, quantum phase transitions and the emergent Fermi-liquid,” arXiv:hep-th/0904.1993.

[12] S. A. Hartnoll, C. P. Herzog and G. T. Horowitz, “Holographic superconductors,” JHEP 0812 92008) 015 [arXiv:hep-th/0810.1563].

[13] S. Sachdev and M. Muller, “Quantum criticality and Black holes”, arXiv:cond-mat.str-el/0810.3005.

[14] S. Sachdev, “Finite temperature dissipation and transport near quantum critical points”, arXiv:cond-mat.str-el/0910.1139.

[15] T. Faulkner, N. Iqbal. H. Liu, J. McGreevy, and D. Vegh, “From black holes to strange metals,” arXiv-hep-th/ 1003.1728.

[16] S. A. Hartnoll, “Lectures on holographic methods for condensed matter physics,” [arXiv:hep-th/0903.3246].

[17] N. Iqbal and H. Liu, “Universality of the hydrodynamic limit in AdS/CFT and the membrane paradigm”, Phys. Rev. D79 (2009) 025023 [arXiv:hep-th/0809.3808].

[18] C. P. Herzog, “Lectures on holographic superfluidity and superconductivity,” Phys. A42 (2009) 343001 [arXiv:hep-th/0904.1975].

[19] S. Sachdev, *Quantum Phase transitions*, Cambridge University Press, Cambridge (1999).

[20] A. S. T. Pires, L. S. Lima and M. E. Gouvea, “The phase diagram and critical properties of the two-dimensional anisotropic XY model”, J. Phys.:Condens. Matter 20 (2008) 015208.

[21] A. S. T. Pires, M. E. Gouvea, “Quantum phase transitions in the two-dimensional XY model with single-ion anisotropy”, Physica A 388(2009)21.

[22] A. Altland and B. Simons, *Condensed Matter Field Theory*, Cambridge University Press, Cambridge (2007).

[23] J. L. Petersen, “Introduction to the Maldacena conjecture on AdS/CFT,” arXiv-hep-th/9902.131.

[24] E. J. Brynjolfsson, U. H. Danielsson, L. Thorlacius, and T. Zingg, “Black hole thermodynamics and heavy fermion metals,” arXiv:hep-th/1003.5361.

[25] L. S. Lima and A. S. T. Pires, “Dynamics of the anisotropic two-dimensional XY model,” Eur. Phys. J. B70 (2009) 335.

[26] A. S. T. Pires and L. S. Lima, “Spin transport in the anisotropic easy-plane tw-dimensional Heisenberg antiferromagnet,” J. Mag. Mag. Mat. 322 (2010) 668.

[27] N. Itzhaki, J. M. Maldacena, J. Sonnenschein and S. Yankielowicz, “Supergravity and the large N limit of theories with sixteen supercharges,” Phys. Rev. D58 (1998) 046004, ArXiv:hep-th/9802042.

[28] S. M. Carrol, *Spacetime and geometry,* Addison- Wesley, USA, 2004.

[29] G. W. Gibbons and S. W. Hawking, “Action integrals and partition functions in quantum gravity,” Phys. Rev. D15 (1977) 2752.

[30] O. Harony, S. S. Gubser, J. Maldacena, H. Oooguri, and Y. Oz, “Large N field theories, string theory and gravity,” arXiv:hep-th/9905111.